\documentclass{article}
\begin{document}
\title{Optical metrics and birefringence of anisotropic media}
\author{
Alexander B. Balakin\footnote{Electronic address: Alexander.Balakin@ksu.ru}\\
Department of General Relativity and Gravitation  \\
Kazan State University, 420008 Kazan, Russia\\and\\
Winfried Zimdahl\footnote{Electronic address:
zimdahl@thp.uni-koeln.de}\\
Institut f\"ur Theoretische Physik, Universit\"at zu  K\"oln\\
D-50937 K\"oln, Germany }
\date{\today}
\maketitle

\begin{abstract}
The material tensor of linear response in electrodynamics is
constructed out of products of two symmetric second rank tensor
fields which in the approximation of geometrical optics and for
uniaxial symmetry reduce to ``optical" metrics, describing the
phenomenon of birefringence. This representation is interpreted in
the context of an underlying internal geometrical structure
according to which the symmetric tensor fields are vectorial
elements of an associated two-dimensional space.
\end{abstract}
\vspace{0.8cm} Key words: Anisotropic media, geometrical optics,
optical metric, birefringence\\
\vspace{0.5cm}
\section{Introduction}

Electrodynamics in media is a broadly investigated subject with a
variety of different facets \cite{LL,Bressan,Mauginbook}. As far
as the propagation of waves is concerned, the approximation of
geometrical optics has played a major role. Within this approach
Maxwell's equations are transformed into a set of algebraic
equations for the wave vector. In a relativistic context (see
\cite{Gordon,PMQ,Ehlers,Perlick}), a specific feature of the wave
propagation in media is the appearance of an ``optical" metric,
introduced by Gordon \cite{Gordon}. With respect to this quantity
light in isotropic media propagates as in vacuum. Light rays
follow geodesics \cite{PMQ} and the wave vector is a null vector
of this metric. The optical metric is characterized by the
refraction index of the medium which is composed of the
coefficients of dielectricity and magnetic permeability. The
latter quantities, in turn, enter the material tensor that relates
the excitation (induction) tensor to the field strength tensor.
For the motivation of this paper it is essential that this happens
in a manner which allows one to represent the fourth rank material
tensor as a combination of products of the second rank optical
metric. Written in terms of the optical metric, the material
tensor for an isotropic medium has the same structure as the
``material" tensor for the vacuum, only that the spacetime metric
within the latter is replaced by the optical metric
\cite{MauginJMP}. Here we ask to what extent this type of
representation can be generalized to anisotropic media. This is
anything but straightforward since light propagation in
anisotropic media is accompanied by the phenomenon of
birefringence. A hint here is the fact that in the simplest
anisotropic case, the case of uniaxial symmetry, birefringence can
be described by two Lorentzian optical metrics \cite{Perlick}
(under more general circumstances they have to be replaced by
Finsler metrics). Given the mentioned representation of the
material tensor in the isotropic case, this raises the question
whether there exists a bi-metric representation of this tensor in
anisotropic media. While the problems of birefringence and
bi-metricity have been considered for non-linear electrodynamics
(birefringence as a non-linear phenomenon) \cite{Vis8,Obukhov1},
we are not aware of an exlicit representation of the material
tensor in terms of two optical metrics in the linear case,
describing the so-called {\it ordinary} and {\it extraordinary}
waves in uniaxial media. We shall demonstrate that the material
tensor can indeed be written as a combination of products of those
two optical metrics which characterize the light cone structure
for uniaxial symmetry. A construction of the material tensor out
of two symmetric second rank tensor fields is even possible in
general anisotropic media, but the interpretation of these
quantities as optical metrics is restricted to the uniaxial case.
We also show that the transition between different representations
of the material tensor in terms of different pairs of second rank
tensor fields is governed by invariance properties in an
associated two-dimensional internal vector space.

The paper is organized as follows. In section \ref{Maxwell's
equations} we consider Maxwell's equations and the general
constitutive laws which mediate between the tensors of excitation
(induction) and field strength. We decompose the material tensor
with respect to contribution parallel and orthogonal to the four
velocity of the medium. We recall the role of an optical metric in
isotropic media and provide different representations of the
material tensor in terms of this metric. Section \ref{Geometrical
optics} is a collection of basic relations of geometrical optics.
Section \ref{Anisotropic media}, which is the main section of the
paper, is devoted to the construction of two symmetric second rank
tensor field in terms of which a representation of the material
tensor is obtained. Novel geometric features of this
representation are pointed out. In the approximation of
geometrical optics Maxwell's equations reduce to a fourth order
algebraic equation for the wavevector, the ``extended" Fresnel
equation (cf. \cite{HehlObukhov}). For the case of uniaxial
symmetry this equation is shown to factorize where the two
symmetric tensor fields reduce to optical metrics for ordinary and
extraordinary wave propagation. A summary of the paper is given in
section \ref{Discussion}.

\section{Maxwell's equations and constitutive laws}
\label{Maxwell's equations}

\subsection{General theory}

The covariant, source free Maxwell equations in continuous
electromagnetic media are
\\
\begin{equation}
\nabla_{k} F^{*ik} =0 \ \quad {\rm and} \quad \nabla_{k} H^{ik} =
0 \ , \label{maxwell}
\end{equation}
\\
where
\\
\begin{equation}
F^{*ik} = \frac{1}{2 \sqrt{-g}}\epsilon^{ikls} F_{ls} \label{dual}
\end{equation}
\\
is the dual to the field strength tensor $F^{ik}$ and $H^{ik}$ is
the induction (or excitation) tensor. Both $F^{ik}$ and $H^{ik}$
are antisymmetric, i.e., $F^{ik} =- F^{ki}$ and $H^{ik}=- H^{ki}$.
Here, $\nabla_{k}$ denotes the covariant derivative with respect
to the spacetime metric $g_{ik}$ (with determinant $g$ and
signature $+---$) and $\frac{1}{\sqrt{-g}}\epsilon^{ikls}$ is the
Levi-Civita (pseudo-) tensor where $\epsilon^{ikls}$ is completely
antisymmetric with $\epsilon^{0123} = - \epsilon_{0123} = 1$.

The set of equations (\ref{maxwell}) has to be completed by
constitutive relations, linking the electromagnetic inductions to
the field strengths. Restricting ourselves to linear dielectric,
permeable media, the most general structure for the corresponding
constitutive laws  is
\\
\begin{equation}
H^{ik} = C^{ikmn}F_{mn} , \label{linlaw}
\end{equation}
\\
where $C^{ikmn}$ is the material tensor which characterizes the
electromagnetic properties
of the medium in the linear response approximation. This tensor has
the symmetries
\\
\begin{equation}
C^{ikmn} = C^{mnik} = - C^{kimn} = - C^{iknm} \ .\label{symmC}
\end{equation}
\\
Consequently, it has 21 independent components. (See, however,
\cite{HehlObukhov} for an approach without the symmetry $C^{ikmn}
= C^{mnik}$). Assuming that the medium is characterized by a
four-velocity field $U^i$, normalized by $U^i U_i =1$, we may
define the vectors $D^i$, $H^i$,$E^i$, and $B^i$ as
\cite{Lichnero}
\\
\begin{equation}
D^i = H^{ik} U_k\ , \quad
H^i = H^{*ik} U_k\ , \quad
E^i = F^{ik} U_k\ , \quad
B^i = F^{*ik} U_k\ .
\label{dheb}
\end{equation}
\\
These vectors are orthogonal to the four-velocity vector $U^i$,
\begin{equation}
D^i U_i = 0 = E^i U_i\ , \quad H^i U_i = 0 = B^i U_i\ .
\label{ortho}
\end{equation}
The field strength tensor can be decomposed according to

\begin{equation}
F_{mn} = \delta^{pq}_{mn}E_p U_q  - \eta_{mnl} B^l\ , \label{Fmn}
\end{equation}
where $\delta^{ik}_{mn}$ is the generalized 4-indices $\delta-$Kronecker
tensor
\[
\delta^{ik}_{mn} =  \delta^{i}_{m}\delta^{k}_{n} -
\delta^{i}_{n}\delta^{k}_{m}\
\]
and $\eta_{mnl}$ is an antisymmetric tensor orthogonal to $U^i$,
defined as
\begin{equation}
\eta_{mnl} \equiv \sqrt{-g}\epsilon_{mnls} U^s,
\quad
\eta^{ikl} \equiv \frac{1}{\sqrt{-g}}\epsilon^{ikls} U_s\ .
\label{eta}
\end{equation}
A useful identity is
\begin{equation}
- \eta^{ikp} \eta_{mnp} = \delta^{ikl}_{mns} U_l U^s =
\Delta^i_m \Delta^k_n - \Delta^i_n \Delta^k_m \ ,
\label{iden1}
\end{equation}
where
\[
\delta^{ikl}_{mns} = \delta^{i}_{m}\delta^{k}_{n}\delta^{l}_{s}
+\delta^{k}_{m}\delta^{l}_{n}\delta^{i}_{s}
+\delta^{l}_{m}\delta^{i}_{n}\delta^{k}_{s} -
\delta^{i}_{m}\delta^{l}_{n}\delta^{k}_{s} -
\delta^{l}_{m}\delta^{k}_{n}\delta^{i}_{s} -
\delta^{k}_{m}\delta^{i}_{n}\delta^{l}_{s}
\]
and
\begin{equation}
\Delta^{ik} = g^{ik} - U^i U^k\ .
\label{proj}
\end{equation}
Upon contraction, equation (\ref{iden1}) yields the further identity
\begin{equation}
\frac{1}{2} \eta^{ikl}  \eta_{klm} = - \delta^{il}_{ms} U_l U^s
= - \Delta^i_m .
\label{iden2}
\end{equation}

\subsection{Decomposition of the material tensor  $C^{ikmn}$}

We can uniquely decompose the tensor $C^{ikmn}$  as
\begin{eqnarray}
C^{ikmn} &=& \frac12 \left[
\varepsilon^{im} U^k U^n - \varepsilon^{in} U^k U^m +
\varepsilon^{kn} U^i U^m - \varepsilon^{km} U^i U^n \right] \nonumber\\
&&-\frac12
\eta^{ikl}(\mu^{-1})_{ls}  \eta^{mns}  \nonumber\\
&& -\frac12 \left[\eta^{ikl}(U^m\nu_{l \ \cdot}^{\ n} - U^n \nu_{l
\ \cdot}^{\ m}) + \eta^{lmn}(U^i \nu_{l \ \cdot}^{\ k} - U^k
\nu_{l \ \cdot}^{\ i} ) \right] . \label{Cdecomposition}
\end{eqnarray}
Here $\varepsilon^{im}$, $(\mu^{-1})_{pq}$ and $\nu_{p \ \cdot}^{\ m}$
are defined as
\begin{eqnarray}
\varepsilon^{im} &=& 2 C^{ikmn} U_k U_n\ , \nonumber\\
(\mu^{-1})_{pq} &=& - \frac{1}{2} \eta_{pik} C^{ikmn} \eta_{mnq}\ , \nonumber\\
\nu_{p \ \cdot}^{\ m} &=& \eta_{pik} C^{ikmn} U_n \ .
\label{emunu}
\end{eqnarray}
The tensors $\varepsilon_{ik}$ and $(\mu^{-1})_{ik}$ are
symmetric, but $\nu_{l \ \cdot}^{\ k}$ is in general
non-symmetric. The dot denotes the position of the second index
when lowered. These three tensors are spacelike, i.e., they are
orthogonal to $U^i$,
\begin{equation}
\varepsilon_{ik} U^k = 0\ , \quad (\mu^{-1})_{ik} U^k = 0\ , \quad
\nu_{l \ \cdot}^{\ k} U^l = 0 = \nu_{l \ \cdot}^{\ k} U_k\ .
\label{ortho2}
\end{equation}
The total trace of the $C^{ikmn}$ tensor is equal to
\begin{equation}
C^{ikmn} g_{im}g_{kn} = \varepsilon^k_k +  ( \mu^{-1})^k_{k}  \ .
\label{spur}
\end{equation}
With the help of the definitions (\ref{linlaw}), (\ref{dheb}) and
the decomposition (\ref{Cdecomposition}), we obtain the linear
laws
\begin{equation}
D^i = \varepsilon^{im} E_m - B^l \nu_{l \ \cdot}^{\ i}\  \quad
{\rm and} \quad H_i = (\mu^{-1})_{im} B^m  + \nu_{i \ \cdot}^{\ m}
E_m  \ . \label{linlawDH}
\end{equation}
Obviously, the tensors
$\varepsilon^{im}$ and
$\mu_{pq}$ are the four-dimensional analogues of the dielectricity
tensor and the magnetic permeability tensor, respectively.
The tensor
$\nu_{p \ \cdot}^{\ m}$ describes
magneto-electric cross effects.
Thus, the 21 components of
$C^{ikmn}$ consist of the 6 components of $\varepsilon^{im}$,
the 6 components of $(\mu^{-1})_{pq}$ and the 9 components of
$\nu_{p\ \cdot}^{\ m}$.
For media without cross effects the number of independent components reduces to 12.

\subsection{Special cases}
\subsubsection{Vacuum}

The vacuum is characterized by
\begin{equation}
H^{ik} = F^{ik} = g^{im} g^{kn} F_{mn} \ , \label{linlawvac}
\end{equation}
which corresponds to a ``material" tensor
\\
\begin{equation}
C_{(\rm vac)}^{ikmn} = \frac{1}{2} \left(g^{im} g^{kn} - g^{in}
g^{km} \right)\ . \label{Ctensorvac}
\end{equation}
\\
Under this condition the quantities (\ref{emunu}) reduce to
\\
\begin{equation}
\varepsilon^{ik} = (\mu^{-1})^{ik} = \Delta^{ik}\ , \quad
\nu_{i \ \cdot}^{\cdot \ k} = 0\ .
\label{emunuvac}
\end{equation}
\\
Alternatively, introducing the latter relations into
Eq.~(\ref{Cdecomposition}), we obtain the special case
(\ref{Ctensorvac}).

\subsubsection{Isotropic medium}

A spatially isotropic medium is characterized by
\\
\begin{equation}
D^i = \varepsilon E^i  \quad {\rm and}\quad H_i = \frac{1}{\mu}
B_i \ , \label{DHiso}
\end{equation}
\\
with two phenomenological scalars, the dielectricity $\varepsilon$
and the permeability $\mu$. The relations (\ref{DHiso}) are the
special case of the general structure (\ref{linlawDH}) for
\\
\begin{equation}
\varepsilon^{ik} = \varepsilon \Delta^{ik}\ , \quad
(\mu^{-1})^{ik} = \frac{1}{\mu} \Delta^{ik}\ , \quad
\nu_{i \ \cdot}^{\ k} = 0 \ .
\label{emunuiso}
\end{equation}
\\
For $\varepsilon$ and $\mu$ equal to unity, we recover the vacuum
case (\ref{emunuvac}). Introducing the expressions
(\ref{emunuiso}) for $\varepsilon^{ik}$, $(\mu^{-1})^{ik}$ and
$\nu_{i \ \cdot}^{\ k}$ into (\ref{Cdecomposition}) one obtains
the material tensor
\\
\begin{eqnarray}
C^{ikmn}_{(\rm isotr)} = \frac{1}{2\mu} \left(g^{im} g^{kn} -
g^{in} g^{km} \right) + \left(\frac{\varepsilon \mu -1}{2\mu}
\right) \left(g^{im} U^k U^n \qquad\qquad
\right.&&\nonumber\\\left.- g^{in} U^k U^m + g^{kn} U^i U^m -
g^{km} U^i U^n \right) .&& \label{Cdecompiso}
\end{eqnarray}
\\
A convenient representation is obtained in terms of the second
rank tensor
\begin{equation}
g^{*ik} = \frac{1}{n^2} \left[g^{ik} + \left( n^2 - 1 \right)
U^{i} U^{k} \right] \equiv \frac{1}{n^2} \Delta^{ik} + U^{i}
U^{k}\ , \label{gstar}
\end{equation}
where $n$ is the refractive index, defined by $n^2 = \varepsilon
\mu$. The inverse tensor is
\begin{equation}
g^{*}_{km} = n^2 \left[g_{km} + \left( \frac{1}{n^2} - 1 \right)
U_{k} U_{m} \right] \equiv n^2 \Delta_{km} + U_{k} U_{m} \ ,
\label{gstarinv}
\end{equation}
such that
\begin{equation}
g^{*ik} g^{*}_{km} = \delta^i_m \ . \label{invers}
\end{equation}
Obviously, we have
\begin{equation}
g^{*ik} U_k = U^i , \quad g^{*ik} U_i U_k = 1 , \quad  g^{*}_{km}
U^m = U_k , \quad g^{*}_{km} U^k U^m = 1 , \label{norm}
\end{equation}
i.e., the tensors $g^{*ik}$ and $g^{*}_{km}$ preserve the norm of
the four-velocity vector. (Note, however, that generally indices
are raised and lowered with the metric tensor $g_{ik}$.)
Furthermore, the relation
\\
\begin{equation}
g^{*ln} = \frac{\varepsilon^{ln}}{\varepsilon n^2}  + U^l U^n \
\label{g*eps}
\end{equation}
\\
for the isotropic case $\varepsilon^{ik} = \varepsilon
\Delta^{ik}$ is valid. With the help of both $g^{ik}$ and
$g^{*ik}$ the tensor (\ref{Cdecompiso}) can be written as
\begin{eqnarray}
C^{ikmn}_{(\rm isotr)} = \frac{n^2}{2 \mu} \left\{\left[g^{im}
g^{kn} - g^{in} g^{km}\right] - \frac{n^2}{(n^2-1)}
\left[(g^{im}-g^{*im})(g^{kn}-g^{*kn})\right.\right.&&\qquad\nonumber\\
\qquad \qquad -\left.\left. (g^{in}-g^{*in})(g^{km}-g^{*km})
\right] \right\} \ , &&\label{Ciso}
\end{eqnarray}
which will be useful for later reference. The most compact and
elegant way, however, is the symmetric vacuum type form (cf.
\cite{MauginJMP})
\begin{equation}
C^{ikmn}_{(\rm isotr)} = \frac{n^4}{2 \mu} \left(g^{*im} g^{*kn} -
g^{*in} g^{*km}  \right) \ . \label{Cisogstar}
\end{equation}
\\
In such a form the material tensor $C^{ikmn}_{(\rm isotr)}$ for
the isotropic medium, up to the factor $\frac{\mu}{n^4}$, has the
structure of (\ref{Ctensorvac}), with the metric $g^{ik}$ replaced
by $g^{*ik}$.

\section{Geometrical optics}
\label{Geometrical optics}

\subsection{Basic relations}

We assume the field strength tensor to be given in terms of a four
potential $A_m$ by
\begin{equation}\label{defFmn}
F_{mn} = \nabla_{m}A_{n} - \nabla_{n}A_{m}\ .
\end{equation}
Geometrical optics is based on the leading-order approximation in
the eikonal derivatives of
\begin{equation}
A_l = a_l e^{i \Phi} , \quad k_m \equiv \nabla_m \Phi , \label{A}
\end{equation}
where $\Phi$ is the phase, $a_l$ is a slowly varying amplitude and
$k_m$ is a wave four-vector. In this approximation    (for complex
quantities always the real part has to be taken)
\begin{equation}\label{Fmn}
F_{mn} = i\left(k_{m}A_{n} - k_{n}A_{m}\right)\
\end{equation}
and  Maxwell's equations reduce to
\begin{equation}
 k_l C^{ilmn} (k_m A_n - k_n A_m) = 0 \ .
\label{Maxgo}
\end{equation}

\subsection{Vacuum}

In the vacuum case Eq.~(\ref{Maxgo}) takes the form
\begin{equation}
k^i(g^{ln} k_l A_n) - A^i (g^{lm} k_l k_m) = 0 \ .
\label{Maxgovac}
\end{equation}
\\
Contraction with $U_{i}$ yields
\\
\begin{equation}
\left(U_{i}k^i\right)\left(g^{ln} k_l A_n\right) =
\left(U_{i}A^i\right) \left(g^{lm} k_l k_m\right)  \ . \label{}
\end{equation}
\\
For the Landau gauge $U_{i}A^i = 0$  and $U_{i}k^i \neq 0$ it
follows that
\\
\begin{equation}
g^{ln}k_l A_n = 0 \ . \label{Lorentzvac}
\end{equation}
\\
While this is identical to the Lorentz gauge, it is {\it not}
imposed as a separate condition in the present context but it is a
consequence of the condition $U_{i}A^i = 0$. With
(\ref{Lorentzvac}) the field equation (\ref{Maxgovac}) admits a
nontrivial solution for the potential $A^i$ if and only if
\begin{equation}
g^{lm}k_l k_m = 0 \ , \label{nullvectorvac}
\end{equation}
i.e., the wave four vector $k^i$ has to be a null vector with
respect to the metric $g^{ik}$. Moreover, the curves to which
$k^{a}$ is a tangent vector are geodesics \cite{Stephani}.

\subsection{Isotropic medium}

With the help of the optical metric (\ref{gstar}) and with
(\ref{Cisogstar})  the Maxwell equations for an isotropic medium
can be written in the compact form
\begin{equation}
\frac{n^4}{2 \mu} \left(g^{*im}k_m g^{*ln}k_l A_n - g^{*in} A_n
g^{*lm} k_l k_m  \right) = 0 \ . \label{Maxiso}
\end{equation}
Proceeding as in the vacuum case, projection along $U_{i}$ leads
to
\\
\begin{equation}
\left(U_{i}k^i\right)\left(g^{*ln} k_l A_n\right) =
\left(U_{i}A^i\right) \left(g^{*lm} k_l k_m\right)  \ . \label{}
\end{equation}
\\
For $U_{i}A^i = 0$ (Landau gauge) and  $U_{i}k^i \neq 0$ we get
\begin{equation}
g^{*ln}k_l A_n  = 0 \ ,\label{Lorentziso}
\end{equation}
which, because of $U_{i}A^i = 0$ coincides with (\ref{Lorentzvac})
and the Maxwell equations (\ref{Maxiso}) simplify to
\begin{equation}
g^{*in} A_n g^{*lm} k_l k_m  = 0\ . \label{}
\end{equation}
This equation admits a nontrivial solution for the potential, if
and only if
\begin{equation}
g^{*lm} k_l k_m = 0 \ , \label{nulliso}
\end{equation}
i.e. if the wave four-vector $k_m$ is a null vector with respect
to $g^{*}_{lm}$, which justifies the name optical metric
\cite{Gordon} for this quantity. It was proven by Pham Mau Quan
\cite{PMQ} (see also \cite{Synge}) that the corresponding light
curves are null geodesics of the metric $g^{*}_{ik}$.

\section{Anisotropic media}
\label{Anisotropic media}

\subsection{Representation of the material tensor}

In the isotropic case it turned out to be possible to find a
metric $g ^{*ik}$ in terms of which the matter tensor has the
quasi-vacuum representation (\ref{Cisogstar}). Moreover, $g
^{*ik}$ is an optical metric. With respect to $g ^{*ik}$ the wave
vector is a null vector. Both for the representation of the matter
tensor and for the light propagation in geometrical optics the
transition from vacuum to an isotropic medium consists in
replacing the spacetime metric by the optical metric $g ^{*ik}$.
It is natural to ask, to what extent these features may be
generalized also to anisotropic media. A useful hint here is the
circumstance that for anisotropic media of the simplest kind, the
uniaxial crystal, there exist {\it two} optical metrics which
account for the phenomenon of birefringence (see, e.g.,
\cite{Perlick}). This seems to suggest a {\it bi-metric}
representation of the matter tensor in such media. The
construction of $C^{ikmn}$ out of two different symmetric second
rank tensor fields is the subject of the present section. While
these tensor fields cannot be expected to be optical metrics in
the general case, for an uniaxial symmetry, however, a
decomposition of $C^{ikmn}$ in terms of two optical metrics will
indeed be possible. With respect to each of them light propagation
proceeds as in vacuum in a similar sense as described by
(\ref{nulliso}) for isotropic media. To the best of our knowledge,
this problem has not been addressed in the literature so far. Our
strategy will be as follows. In a first step we simply {\it
assume} the existence of two symmetric tensor fields as
construction elements of the material tensor. This choice is not
unique and we have to single out a suitable pair of symmetric
tensors. This will be done with the help of a geometric
interpretation according to which these tensors can be regarded as
vectors in a two-dimensional space. The next step will then be the
explicit construction of these tensor fields and, finally, the
demonstration that they reduce to two different optical metrics in
the appropriate limit.

   This strategy implies that we consider media for which the
tensor $C^{ikmn}$ in (\ref{Cdecomposition}) is reducible to the
form
\begin{equation}
C^{ikmn} = \frac{1}{2\hat{\mu}}\sum_{(\alpha)(\beta)}
 G_{(\alpha)(\beta)}\left(g^{im (\alpha)} \ g^{kn
(\beta)} - g^{in (\alpha)} \ g^{km (\beta)} \right) ,
\label{supergeneral}
\end{equation}
where $g^{im(a)}$ and $g^{im(b)}$ are a pair of (for the moment
unknown) symmetric, non-degenerate tensor fields. The indices
$\alpha$ and $\beta$ take the values $a$ and $b$ and the sum is
over all possible combinations. By construction, the expression
(\ref{supergeneral}) satisfies the symmetry conditions
(\ref{symmC}). For a given tensor $C^{ikmn}$ with generally 21
independent components the representation (\ref{supergeneral}) may
be regarded as a set of algebraic equations to determine the 10
components of $g^{im (a)}$, the 10 components of $g^{im (b)}$ and,
additionally, one of the three quantities $G_{(a)(a)}/\hat{\mu}$,
$G_{(b)(b)}/\hat{\mu}$, and $G_{(a)(b)}/\hat{\mu}$ =
$G_{(b)(a)}/\hat{\mu}$ (we assume $ G_{(\alpha)(\beta)}$ to be
symmetric). Two of these quantities remain arbitrary, the
corresponding freedom will be used later on. The representation
(\ref{supergeneral}) is a direct generalization of
(\ref{Ctensorvac}) and (\ref{Cisogstar}) for the cases of vacuum
and isotropic media, respectively.   Since we require for the
isotropic limit $g^{im (a)} = g^{im (b)} = g^{*im}$ the
decomposition (\ref{Cisogstar}) with $\hat{\mu} = \frac{\mu}{n^4}$
to hold, we obtain
\\
\begin{equation}
\sum_{(\alpha)(\beta)} G_{(\alpha)(\beta)} = 1 \ \label{sum=1}
\end{equation}
\\
as an additional condition. The choice of the symmetric tensors
$g^{im(a)}$ and $g^{im(b)}$ in the decomposition
(\ref{supergeneral}) is not unique. A first guess could be (cf.
\cite{HehlObukhov}, (D.1.80))
\\
\begin{equation}
C^{ikmn}_{(1,2)} = \frac{1}{4 \hat{\mu}} \left(g^{im (1)} \ g^{kn
(2)} - g^{in (1)} \ g^{km (2)} + g^{im (2)}  \ g^{kn (1)} - g^{in
(2)} \ g^{km (1)} \right)\,, \label{general}
\end{equation}
\\
which corresponds to
\\
\begin{equation}
G_{(1)(1)} = G_{(2)(2)} = 0 , \quad G_{(1)(2)} = G_{(2)(1)} =
\frac{1}{2} , \quad {\rm det}(G_{(1,2)})  = -\frac{1}{4}
\label{sG1}
\end{equation}
\\
where ${\rm det}(G_{(1,2)}) $ is the determinant $G_{(1)(1)}
G_{(2)(2)} - G_{(1)(2)} G_{(2)(1)}$ of the matrix
$G_{(\alpha)(\beta)}$ for this case. So far we will not use,
however a specific choice but further consider the general
representation (\ref{supergeneral}).   As any symmetric tensor,
the quantities $g^{ik(\alpha)}$ can be decomposed with respect to
their components parallel and orthogonal to the four-velocity
$U^{i}$,
\\
\begin{equation}
g^{ik (\alpha)} = {\cal B}^{(\alpha)} U^i U^k + \tilde{{\cal
D}}^{i (\alpha)} U^k + \tilde{{\cal D}}^{k (\alpha)} U^i +
\tilde{{\cal S}}^{ik (\alpha)} \,, \label{gviaU0}
\end{equation}
\\
where
\\
\begin{equation}
{\cal B}^{(\alpha)} \equiv g^{ik (\alpha)} U_i U_k \,, \quad
\tilde{{\cal D}}^{p (\alpha)} \equiv \Delta^p_i g^{ik(\alpha)} U_k
\,, \qquad \tilde{{\cal S}}^{pq (\alpha)} \equiv \Delta^p_i
g^{ik(\alpha)}\Delta^q_k \,. \label{gviaUD0}
\end{equation}
\\
In case $U^{i}$ is lightlike with respect to one or both tensors
$g^{im(\alpha)}$, the corresponding coefficients ${\cal
B}^{(\alpha)} $ vanish. We shall assume, however, that,
analogously to the vacuum and isotropic media cases, the four
velocity $U^{i}$ is timelike with respect to both tensors
$g^{im(\alpha)}$. Under this condition the factors ${\cal
B}^{(\alpha)}$ in front of $U^i U^k$ in (\ref{gviaU0}) are
different from zero and can be absorbed into the coefficients
$G_{(\alpha)(\beta)}$. The decomposition (\ref{gviaU0}) then
reduces to
\\
\begin{equation}
g^{ik (\alpha)} =  U^i U^k + {\cal D}^{i (\alpha)} U^k + {\cal
D}^{k (\alpha)} U^i + {\cal S}^{ik (\alpha)} \,,
\label{gviaU}
\end{equation}
\\
with quantities ${\cal D}^{p (\alpha)}$ and ${\cal S}^{pq
(\alpha)}$ redefined accordingly.
\\
The magneto-electric cross terms $\nu^{\ \ p}_{m \cdot}$ in
(\ref{emunu}) are determined by the coefficients ${\cal D}^{p
(\alpha)}$ via
\begin{equation}
\nu^{\ \ p}_{m \cdot} = \frac{1}{2\hat{\mu}}\sum_{(\alpha)(\beta)}
G_{(\alpha)(\beta)} \eta_{pik} \left( {\cal D}^{k (\beta)} g^{im
(\alpha)} - {\cal D}^{i (\alpha)} \ g^{km (\beta)} \right) \,.
\label{nuD}
\end{equation}
Obviously, a sufficient condition for excluding magneto-electric
cross effects is ${\cal D}^{p (\alpha)} = 0$.   (Note that even in
the presence of cross effects one of the ${\cal D}^{p (\alpha)}$
can be transformed to zero by assuming $U^{i}$ to be an
eigenvector of the corresponding $g^{ik (\alpha)}$). In the
following we shall restrict ourselves to ${\cal D}^{p (\alpha)} =
0$ and use
\\
\begin{equation}
g^{ik (\alpha)} =  U^i U^k + {\cal S}^{ik (\alpha)} \quad
\Rightarrow \quad U_ig^{im (\alpha)} = U^{m} \,. \label{gviaU1}
\end{equation}
\\
Applying the structure (\ref{gviaU1}) to (\ref{supergeneral}), the
spatial tensors $\varepsilon^{im}$ and $(\mu^{-1})_{pq}$ in
(\ref{emunu}) can be written as
\\
\begin{equation}
\varepsilon^{im} = \frac{1}{\hat{\mu}} \left\{ \left[G_{(a)(a)} +
G_{(a)(b)} \right]{\cal S}^{im (a)} + \left[G_{(a)(b)} +
G_{(b)(b)} \right]{\cal S}^{im (b)}  \right\} \,, \label{1eviagg}
\end{equation}
\\
and
\\
\begin{eqnarray}
(\mu^{-1})^{pq} &=&  - \frac{1}{\hat{\mu}} \eta^p_{\cdot ik} \
\eta^q_{\cdot mn} \left[ G_{(a)(a)} {\cal S}^{im (a)}{\cal S}^{kn
(a)} + G_{(b)(b)}{\cal S}^{im (b)}{\cal S}^{kn (b)}  \right.
\nonumber\\
&& \qquad\qquad\qquad\ + \left. 2 G_{(a)(b)}{\cal S}^{im (a)}{\cal
S}^{kn (b)} \right] \,, \label{1muviagg}
\end{eqnarray}
\\
respectively.   In a next step we make use of the remaining two
degrees of freedom, mentioned in the discussion following equation
(\ref{supergeneral}).   From (\ref{1eviagg}) it is obvious that
there exists a special configuration, for which $\varepsilon^{im}$
depends only on one of the spatially projected (cf. (\ref{gviaU}))
metrics, say on ${\cal S}^{im (a)}$. This configuration, which we
denote by $(a) = (A)$ and $(b) = (B)$, is characterized by the two
relations
\\
\begin{equation}
G_{(A)(A)} + G_{(B)(A)} = 1\ ,\quad G_{(A)(B)} + G_{(B)(B)} = 0
 \, \label{ConditionsG}
\end{equation}
\\
which obviously satisfy (\ref{sum=1}). Introducing the
abbreviation $\gamma \equiv G_{(A)(B)} = G_{(B)(A)} = - {\rm
det}(G_{(A,B)})$, where the latter quantity denotes the
determinant $G_{(A)(A)} G_{(B)(B)} - G_{(A)(B)} G_{(B)(A)}$ of the
matrix $G_{(A)(B)}$ for this configuration, the material tensor
(\ref{supergeneral}) takes the form
\\
\begin{eqnarray}
C^{ikmn}_{(A,B)} = \frac{1}{2 \hat{\mu}} \left\{ \left[g^{im (A)}
g^{kn (A)} - g^{in (A)} g^{km (A)}\right]
\right.\qquad\qquad\qquad\qquad\qquad\qquad&&\nonumber\\ - \gamma
\left[(g^{im
(A)}-g^{im (B)})(g^{kn (A)}-g^{kn (B)}) \right.\qquad\quad&&\nonumber\\
- \left.\left. (g^{in (A)}-g^{in (B)})(g^{km (A)}-g^{km (B)})
\right] \right\} \,,&& \label{maindecomp}
\end{eqnarray}
\\
which generalizes the representation (\ref{Ciso}) for the
isotropic case.    The factor $\gamma$ generalizes the (positive)
expression $\frac{n^{2}}{n^{2} - 1}$ in (\ref{Ciso}) and has to be
positive as well (cf. subsection \ref{geometric} below).   The
quantities  (\ref{1eviagg}) and (\ref{1muviagg}) specify to
\begin{equation}
\varepsilon^{im} = \frac{1}{\hat{\mu}}  {\cal S}^{im (A)} \,,
\label{2eviagg}
\end{equation}
and
\begin{eqnarray}
(\mu^{-1})^{pq} &=&  {-} \frac{1}{\hat{\mu}} \eta^p_{\cdot ik} \
\eta^q_{\cdot mn} \left[ {\cal S}^{im (A)}{\cal S}^{kn
(A)}\right. \qquad\qquad\qquad\qquad\qquad\qquad\qquad\nonumber\\
&&\qquad\qquad\qquad - \left. \gamma \left( {\cal S}^{im (A)} {-}
{\cal S}^{im (B)}\right) \left( {\cal S}^{kn (A)} {-} {\cal S}^{kn
(B)}\right) \right] \,, \label{2muviagg}
\end{eqnarray}
respectively.

It is evident from (\ref{gviaU1}) that  (\ref{2eviagg}) implies
\begin{equation}
\varepsilon^{im} = \frac{1}{\hat{\mu}}\left[g^{im (A)} - U^i U^m
\right] \,, \label{eviagg1}
\end{equation}
independent of the value of the parameter $\gamma$. This parameter
contributes to the tensor $(\mu^{-1})^{pq}$ only. In turn,
(\ref{eviagg1}) provides us with an explicit expression for $g^{im
(A)}$ in terms of the tensor $\varepsilon^{im}$, which in the
isotropic limit reduces to the optical metric (\ref{g*eps}).

\subsection{Geometric implications}
\label{geometric}

The change between the explicit representations (\ref{general}) to
(\ref{maindecomp}) of the material tensor corresponds to a
transformation    ($\gamma > 0$)
\\
\begin{equation}
g^{ik(1)} = g^{ik(A)} + \sqrt{\gamma} \left( g^{ik(A)} - g^{ik(B)}
\right)  \label{trafo1AB}
\end{equation}
\\
and
\\
\begin{equation}
g^{ik(2)} = g^{ik(A)} - \sqrt{\gamma} \left(g^{ik(A)} -
g^{ik(B)}\right) , \label{trafo2AB}
\end{equation}
\\
or, inversely,
\\
\begin{equation}
g^{ik (A)}  =  \frac{1}{2} \left( g^{ik (1)} + g^{ik (2)} \right)
\label{trafoA12}
\end{equation}
\\
and
\\
\begin{equation}
g^{ik(B)} = g^{ik(A)}+ \frac{1}{2\sqrt{\gamma}} \left( g^{ik(2)} -
g^{ik(1)} \right) . \label{trafoB12}
\end{equation}
\\
Transformations between different sets of symmetric tensors in the
representation (\ref{supergeneral}) can be regarded as
transformations in a two-dimensional space with a metric
$G_{(\alpha)(\beta)}$, which is symmetric and non-degenerate,
i.e.,
\\
\begin{equation}
G_{(\alpha)(\beta)} = G_{(\beta)(\alpha)} , \qquad G \equiv {\rm
det}(G_{(\alpha)(\beta)}) \neq 0 \,. \label{Gsymm}
\end{equation}
\\
This geometric interpretation requires a Kronecker delta
$\delta^{(\alpha)}_{(\beta)}$ to exist together with a
contravariant tensor $G^{(\mu)(\nu)}$, satisfying
$G^{(\mu)(\nu)}G_{(\nu)(\beta)} = \delta^{(\mu)}_{(\beta)}$.
Furthermore, in such a space there exists an anti-symmetric
Levi-Civita (pseudo-) tensor
\\
\begin{equation}
E_{(\alpha)(\beta)} = \sqrt{\vert G \vert} \
\epsilon_{(\alpha)(\beta)} , \quad \epsilon_{(\alpha)(\beta)} = -
\epsilon_{(\beta)(\alpha)} , \quad \epsilon_{(1)(2)} = 1 .
\label{levi1}
\end{equation}
\\
The point here is that the decomposition (\ref{supergeneral}) is
invariant with respect to transformations in this associated
space. In order to demonstrate this, we consider linear
transformations
\\
\begin{equation}
g^{mn(\alpha)^{\prime}} = t^{(\alpha)^{\prime}} _{(\alpha)}
g^{mn(\alpha)} , \quad G_{(\alpha)^{\prime}(\beta)^{\prime}} =
t^{(\alpha)} _{(\alpha)^{\prime}} t^{(\beta)} _{(\beta)^{\prime}}
G_{(\alpha) (\beta)} , \label{lintrafo}
\end{equation}
\\
and require that the scalar product $G_{(\alpha)
(\beta)}g^{mn(\alpha)}g^{ik(\beta)}$ remains invariant, i.e.,
\\
\begin{equation}
G_{(\alpha)^{\prime} (\beta)^{\prime}}g^{im(\alpha)^{\prime}}
g^{ln(\beta)^{\prime}}  = G_{(\alpha) (\beta)}g^{im(\alpha)}
g^{ln(\beta)}  . \label{invG}
\end{equation}
\\
Contraction with $U_i U_m$ yields
\\
\begin{equation}
\sum _{(\alpha)^{\prime}(\beta)^{\prime}}G_{(\alpha)^{\prime}
(\beta)^{\prime}}g^{ln(\beta)^{\prime}}  = \sum _{(\alpha)
(\beta)}G_{(\alpha) (\beta)}g^{ln(\beta)}  . \label{contractinvG}
\end{equation}
\\
Identifying now $(\alpha)^{\prime}$ and $(\beta)^{\prime}$ with
$(A)$ and $(B)$, respectively, as well as $(\alpha)$ and $(\beta)$
with $(1)$ and $(2)$, respectively, upon using (\ref{ConditionsG})
and (\ref{sG1}) we reproduce the transformation (\ref{trafoA12}).
Similarly, the invariance
\\
\begin{equation}
E_{(\alpha)^{\prime} (\beta)^{\prime}}g^{im(\alpha)^{\prime}}
g^{ln(\beta)^{\prime}}  = E_{(\alpha) (\beta)}g^{im(\alpha)}
g^{ln(\beta)}   \label{invE}
\end{equation}
\\
via
\\
\begin{equation}
\sum _{(\alpha)^{\prime} (\beta)^{\prime}}E_{(\alpha)^{\prime}
(\beta)^{\prime}}g^{ln(\beta)^{\prime}}  = \sum _{(\alpha)
(\beta)}E_{(\alpha) (\beta)}g^{ln(\beta)}  , \label{}
\end{equation}
\\
provides us with the second transformation (\ref{trafoB12}). As a
consistency check, we also realize that
\\
\begin{equation}
G^{\prime} =\vert \frac{\partial g}{\partial g^{\prime}}\vert^{2}
G= 4 \gamma G , \label{determinants}
\end{equation}
\\
which is the correct relations between the determinants
$G^{\prime} \equiv {\rm det}(G_{(A,B)})$ and $G \equiv {\rm
det}(G_{(1,2)})$. Here, $g$ in (\ref{determinants}) stands
symbolically for the set $g^{ik(1)}$, $g^{ik(2)}$, and
$g^{\prime}$ stands for the set $g^{ik(A)}$, $g^{ik(B)}$.

The relations (\ref{trafo1AB}) - (\ref{determinants}) constitute
an underlying geometric structure for the representation of the
material tensor in terms of second rank symmetric tensor fields.
The coefficients $G_{(\alpha) (\beta)}$ play the role of a metric
in the associated two-dimensional space, the tensors
$g^{ik(\alpha)}$ are vectorial objects in this space.
\ \\

\subsection{Maxwell's equations}

With the general expression (\ref{supergeneral}) Maxwell's
equations are
\\
\begin{eqnarray}
C^{ilmn} k_l k_m A_n = 0 \qquad\qquad\qquad\qquad\qquad\qquad\qquad
\qquad\qquad\qquad\qquad &&\nonumber\\
\Rightarrow \frac{1}{2\hat{\mu}}\sum_{(\alpha)(\beta)}
G_{(\alpha)(\beta)}\left(g^{im (\alpha)} \ g^{ln (\beta)} - g^{in
(\alpha)} \ g^{lm (\beta)} \right) k_l k_m A_n  = 0 .
\label{gosuper1}
\end{eqnarray}
\\
Contracting these equations with $U_i$ similarly as for the
previous isotropic case,  we obtain
\\
\begin{equation}
U^mk_m\sum_{(\alpha)(\beta)} G_{(\alpha)(\beta)}g^{ln (\beta)}k_l
 A_n  = U^{n}  A_n \ \sum_{(\alpha)(\beta)}g^{lm (\beta)}  k_lk_m
\ .  \label{}
\end{equation}
\\
With the Landau-gauge $U^{n}  A_n = 0$ and $U^{n}k_n \neq 0$, the
relation generalizing (\ref{Lorentziso}) is
\\
\begin{equation}
g^{ln}_{({\small\rm eff})}k_l
 A_n  = 0\ ,\quad g^{ln}_{({\small\rm eff})} \equiv \sum_{(\alpha)(\beta)}
 G_{(\alpha)(\beta)}g^{ln (\beta)}
\ .  \label{Lorentzaniso}
\end{equation}
\\
The structure of this relation suggests to identify
$g^{ln}_{({\small\rm eff})}$ with one of the symmetric tensors
$g^{ln(\alpha)}$, let's say $g^{ln(a)}$, out of which the material
tensor is constructed. This is equivalent to
\\
\begin{eqnarray}
 &&g^{ln}_{({\small\rm eff})} = g^{ln(a)} =
 \left[G_{(a)(a)} + G_{(b)(a)} \right]g^{ln (a)} +
 \left[G_{(a)(b)} + G_{(b)(b)} \right]g^{ln (b)} \nonumber\\
 && \qquad\quad\Rightarrow \quad G_{(a)(a)} + G_{(b)(a)} = 1 , \qquad
 G_{(a)(b)} + G_{(b)(b)} = 0
\ .  \label{identification}
\end{eqnarray}
\\
It is remarkable, that this choice again results in the previously
derived set of coefficients, characterized by (\ref{ConditionsG}).
The identification $g^{ln}_{({\small\rm eff})} = g^{ln(a)}$ in
(\ref{identification}) is an alternative way to single out this
option which we again denote by $(A)$ and $(B)$. In the present
context we have the additional relation
\\
\begin{equation}
g^{ln (A)}k_l A_n  = 0\ ,\ \label{LorentzgA}
\end{equation}
\\
which has the structure of a generalized Lorentz gauge. We
emphasize again (see the remarks following (\ref{Lorentzvac}))
that this is not imposed as a condition here but that it follows
from the Landau gauge $U^{n} A_n = 0$.

\subsection{Constructing $g^{kn(A)}$ and $g^{kn(B)}$}
\subsubsection{General relations}

While the tensor $g^{ik(A)}$ is given in terms of
$\varepsilon^{ik}$ by (\ref{eviagg1}), it is less straightforward
to obtain a corresponding expression for $g^{ik(B)}$. Following
\cite{Perlick} we consider the eigenvectors $X^i_{(1)},X^i_{(2)}$,
and $X^i_{(3)}$  of the tensor $\varepsilon^{ik}$. These
eigenvectors have the properties
\begin{equation}
g_{ik} X^i_{(a)} X^k_{(b)} = - \delta_{(a)(b)} , \quad
g_{ik} X^i_{(a)} U^k = 0 ,
\label{eigen1}
\end{equation}
\begin{equation}
X^i_{(1)}X^k_{(1)} + X^i_{(2)}X^k_{(2)} + X^i_{(3)}X^k_{(3)} = -
g^{ik} + U^i U^k,
\label{eigen2}
\end{equation}
\begin{equation}
\eta^{i}_{kl} X^k_{(a)}X^l_{(b)} = \epsilon_{(a)(b)(c)} X^i_{(c)},
\label{eigen21}
\end{equation}
\\
where $(a), (b).... = (1), (2), (3)$ are tetrad indices and the
double index $(c)$ denotes a sum. Relation (\ref{eigen21}) is a
consequence of the fact that the scalar function $ \eta_{ikl}
X^i_{(a)} X^k_{(b)} X^l_{(c)}$ is totally antisymmetric with
respect to indices $(a),(b),(c)$ and has to be proportional to the
Levi-Civita symbol $\epsilon_{(a)(b)(c)}$.  The dielectric
permeability can be decomposed according to
\begin{equation}
\varepsilon^{ik} = - \sum_{(a)}\varepsilon_{(a)}  X^i_{(a)}
X^k_{(a)}, \quad \varepsilon^{k}_k =  \varepsilon_{(1)} +
\varepsilon_{(2)} + \varepsilon_{(3)}, \label{exx}
\end{equation}
where the terms $\varepsilon_{(a)}$ denote the    (positive)
eigenvalues, corresponding to the eigenvector $X^i_{(a)}$. Since
the tensor $\varepsilon^{ik}$ is orthogonal to the four-velocity
vector $U^i$, the corresponding eigenvalue $\varepsilon_{(0)}$ is
equal to zero, and the velocity does not appear in this
decomposition.

The  tetrad components  $k_{(a)}$ and $A_{(b)}$  of the wave
four-vector and of the electromagnetic potential four-vector,
respectively, are
\begin{equation}
k_{(a)} \equiv k_i X^i_{(a)} , \quad  A_{(b)} \equiv A_i X^i_{(a)}
\  \label{katetradef}
\end{equation}
and we may write
\begin{equation}
 k^i = U^i (U^l k_l) - k_{(a)} X^i_{(a)} ,
\quad  A^i = - A_{(a)} X^i_{(a)} . \label{decompkA}
\end{equation}
\ \\
The  tensor fields $g^{im (\alpha)}$ are describable in terms of
$U^i$ and $X^i_{(a)}$ according to (cf. (\ref{gviaU1}))
\\
\begin{equation}
g^{im (\alpha)} = U^i U^m + {\cal S}^{(a)(b)}_{(\alpha)} X^i_{(a)}
X^m_{(b)}  \ ,\label{ansatzgalpha}
\end{equation}
\\
where
\\
\begin{equation}
{\cal S}^{(a)(b)}_{(\alpha)}  = {\cal
S}^{mn}_{(\alpha)}X^{(a)}_{m} X^{(b)}_{n}   \ ,\label{Stetrad}
\end{equation}
\\
and where we again have neglected the magneto-electric cross
effects. In the following we use the first and second equations
(\ref{emunu}) to construct the tensors $g^{im (A)}$ and $g^{im
(B)}$.

\subsubsection{The tensor $g^{im (A)}$}

While $g^{im (A)}$ is already known to be given by
(\ref{eviagg1}), we briefly sketch how it is alternatively derived
in the tetrad formalism. Together with (\ref{ansatzgalpha}) the
first equation (\ref{emunu}) becomes
\begin{equation}
2 C^{ikmn} U_k U_n = - \varepsilon_{(a)}  \delta^{(a)(b)}
X^i_{(a)} X^m_{(b)}. \label{ehat}
\end{equation}
With $\Delta^{ik} = - \delta^{(a)(b)} X^i_{(a)} X^k_{(b)}$ it
takes the form
\begin{equation}
\frac{1}{\hat{\mu}} X^i_{(a)} X^m_{(b)} \sum_{(\alpha)(\beta)}
G_{(\alpha)(\beta)} {\cal S}^{(a)(b)}_{(\alpha)} = -
\varepsilon_{(a)}  \delta^{(a)(b)} X^i_{(a)} X^m_{(b)} .
\label{ehat1}
\end{equation}
Contraction with the tetrad vectors yields
\begin{equation}
\sum_{(\alpha)(\beta)} G_{(\alpha)(\beta)} {\cal
S}^{(a)(b)}_{(\alpha)} =
 - \hat{\mu} \varepsilon_{(a)}  \delta^{(a)(b)} \ .
\label{eq1}
\end{equation}
With the coefficients (\ref{ConditionsG}) we obtain
\begin{equation}
{\cal S}^{(a)(b)}_{(A)} = - \hat{\mu} \varepsilon_{(a)}
\delta^{(a)(b)} , \label{eq1A}
\end{equation}
such that the first of the tensors is
\begin{equation}
g^{im (A)} = U^i U^m + \hat{\mu} \varepsilon^{im} \ .
\label{gAgeneral}
\end{equation}
This reproduces (\ref{eviagg1}).  For the choice $\hat{\mu} =
\frac{1}{\mu \varepsilon^2}$ and for $\varepsilon^{ik} =
\varepsilon \Delta^{ik}$ we recover the
optical metric (\ref{g*eps}) for the isotropic medium.\\

\subsubsection{The tensor $g^{im (B)}$}

Applying the tetrad representation (\ref{ansatzgalpha}) to  the
second equation (\ref{emunu}) yields
\begin{equation}
- \frac{1}{2} \eta_{pik} C^{ikmn} \eta_{mnq} = -
(\mu^{-1})_{(a)(b)} X^{(a)}_{p} X^{(b)}_{q} , \label{muhat}
\end{equation}
where $- (\mu^{-1})_{(a)(b)}$ are the tetrad components of the
magnetic permeability tensor. With (\ref{eigen21}) it transforms
into
\begin{equation}
G_{(\alpha)(\beta)} {\cal S}^{(c)(d)}_{(\alpha)}{\cal S}^{(e)(f)}
_{(\beta)} \epsilon_{(a)(c)(e)} \epsilon_{(b)(d)(f)} = 2\hat{\mu}
(\mu^{-1})_{(a)(b)} . \label{eq2}
\end{equation}
Here $\epsilon_{(a)(c)(e)}$ is the Levi-Civita symbol in
three-dimensional space, tetrad indices being regarded as
Euclidean. The equations (\ref{eq2}) form a system of 6 quadratic
equations for the 6 components of ${\cal S}^{(a)(b)}_{(B)}$. Let
us restrict ourselves to the case of a diagonal
$(\mu^{-1})_{(a)(b)}$ tensor, i.e., to
\begin{equation}
(\mu^{-1})_{(a)(b)} = \frac{1}{\mu_{(a)}} \delta_{(a)(b)}\,,\qquad
\mu_{(a)} > 0   \, . \label{mudiag}
\end{equation}
This ansatz allows us to search for an explicit solution of the
basic equations (\ref{eq2}) for which  the ${\cal
S}^{(a)(b)}_{(B)}$ are diagonal, i.e.,
\begin{equation}
{\cal S}^{(a)(b)}_{(B)} = 0 , \ \  {\rm if}\ \  (a) \neq (b) .
\label{uni7}
\end{equation}
Using the structures (\ref{ConditionsG}) of $G_{(\alpha)(\beta)}$
and (\ref{mudiag}) of $(\mu^{-1})_{(a)(b)}$, we obtain from
(\ref{eq2}) the system of equations
\begin{equation}
{\cal S}^{(2)(2)}_{(B)} {\cal S}^{(3)(3)}_{(B)} + {\cal
S}^{(2)(2)}_{(B)} \hat{\mu} \varepsilon_{(3)} + {\cal
S}^{(3)(3)}_{(B)} \hat{\mu} \varepsilon_{(2)} + \left( 1 -
\gamma^{-1} \right) \hat{\mu}^2 \varepsilon_{(2)}
\varepsilon_{(3)} = -\frac{\hat{\mu}}{\gamma\mu_{(1)} }
,\label{diag1}
\end{equation}
\begin{equation}
{\cal S}^{(1)(1)}_{(B)} {\cal S}^{(3)(3)}_{(B)} + {\cal
S}^{(1)(1)}_{(B)} \hat{\mu} \varepsilon_{(3)} + {\cal
S}^{(3)(3)}_{(B)} \hat{\mu} \varepsilon_{(1)} + \left( 1 -
\gamma^{-1}\right) \hat{\mu}^2 \varepsilon_{(1)} \varepsilon_{(3)}
= - \frac{\hat{\mu}}{\gamma\mu_{(2)}} , \label{diag2}
\end{equation}
\begin{equation}
{\cal S}^{(1)(1)}_{(B)} {\cal S}^{(2)(2)}_{(B)} + {\cal
S}^{(1)(1)}_{(B)} \hat{\mu} \varepsilon_{(2)} + {\cal
S}^{(2)(2)}_{(B)} \hat{\mu} \varepsilon_{(1)} + \left( 1 -
\gamma^{-1}\right) \hat{\mu}^2 \varepsilon_{(1)} \varepsilon_{(2)}
= -\frac{\hat{\mu}}{\gamma\mu_{(3)}} .\label{diag3}
\end{equation}
\\
After the substitution
\begin{equation}
{\cal S}^{(a)(b)}_{(B)} = - \hat{\mu} \varepsilon_{(a)}
\delta^{(a)(b)} + Y^{(a)(b)} ,\label{substitution}
\end{equation}
where the first term on the right-hand side coincides with ${\cal
S}^{(a)(b)}_{(A)}$ from (\ref{eq1A}), the system
(\ref{diag1})-(\ref{diag3}) takes the form
\begin{equation}
Y^{(2)(2)} Y^{(3)(3)} = \frac{\hat{\mu}^2}{\gamma}
\varepsilon_{(2)} \varepsilon_{(3)} M_{(1)} , \label{diag1Y}
\end{equation}
\begin{equation}
Y^{(1)(1)} Y^{(3)(3)} = \frac{\hat{\mu}^2}{\gamma}
\varepsilon_{(1)} \varepsilon_{(3)} M_{(2)} , \label{diag2Y}
\end{equation}
\begin{equation}
Y^{(1)(1)} Y^{(2)(2)} = \frac{\hat{\mu}^2}{\gamma}
\varepsilon_{(1)} \varepsilon_{(2)} M_{(3)} , \label{diag3Y}
\end{equation}
where
\begin{equation}
M_{(a)} \equiv 1 - \frac{\varepsilon_{(a)}}{\mu_{(a)}
\hat{\mu}\varepsilon_{(1)}\varepsilon_{(2)} \varepsilon_{(3)} } \
. \label{Ma}
\end{equation}
In the isotropic limit with $\varepsilon_{(a)} = \varepsilon$,
$\mu_{(a)} = \mu$ and $\hat{\mu} = 1/(\varepsilon n^{2})$, the
$M_{(a)}$ quantities vanish identically. Solving the symmetric
system (\ref{diag1Y})-(\ref{diag3Y}) leads to the following
solutions of the system (\ref{diag1})-(\ref{diag3}):
\begin{equation}
{\cal S}^{(1)(1)}_{(B)} = - \hat{\mu} \varepsilon_{(1)} \left[ 1
\pm  \sqrt{\frac{M_{(2)}M_{(3)}}{\gamma M_{(1)}}} \right] ,
\label{co11}
\end{equation}
\begin{equation}
{\cal S}^{(2)(2)}_{(B)} = - \hat{\mu} \varepsilon_{(2)} \left[ 1
\pm  \sqrt{\frac{M_{(1)}M_{(3)}}{\gamma M_{(2)}}} \right] ,
\label{co22}
\end{equation}
\begin{equation}
{\cal S}^{(3)(3)}_{(B)} = - \hat{\mu} \varepsilon_{(3)} \left[ 1
\pm \sqrt{\frac{M_{(1)}M_{(2)}}{\gamma M_{(3)}}} \right] .
\label{co33}
\end{equation}
Thus, the second symmetric tensor is
\begin{equation}
g^{im (B)} = U^i U^m + \hat{\mu} \varepsilon^{im} \mp \hat{\mu}
\sqrt{\frac{M_{(1)} M_{(2)} M_{(3)}}{\gamma}} \sum_{(a)= (1)
}^{(3)} \frac{\varepsilon_{(a)}}{M_{(a)}} X^i_{(a)} X^m_{(a)}\ .
\label{gBgeneral}
\end{equation}
The first two terms on the right-hand side coincide with $g^{im
(A)}$ from (\ref{gAgeneral}). In the isotropic limit we have
$g^{im (B)} = g^{im (A)} = g^{*im}$.  The parameter $\gamma$
remains undetermined. Using the explicit expressions
(\ref{gAgeneral}) and (\ref{gBgeneral}) in (\ref{maindecomp}) it
is obvious that $C^{ikmn}$ is independent of $\gamma$. A proper
choice of $\gamma$ will be essential, however, to make $g^{im
(B)}$ an optical metric in special cases to be discussed below.

\subsection{The characteristic equation}

To obtain the solution of  Maxwell's equations
$C^{ikmn}k_{k}k_{m}A_{n} = 0$ with (\ref{maindecomp}) in the
approximation of geometrical optics, we project these equations
onto the directions $X_{i (a)}$,
\begin{eqnarray}
\gamma\,\left( {\cal S}^{(a)(b)}_{(A)} - {\cal
S}^{(a)(b)}_{(B)}\right) \left[k_{(b)} (k,A)_{(B)} +
A_{(b)}(H_{(A)} - H_{(B)}) \right]\qquad\qquad &&\nonumber\\
-{\cal S}^{(a)(b)}_{(A)} A_{(b)} H_{(A)}  = 0 . \label{CA-CB}
\end{eqnarray}
Here we introduced the following abbreviations:
\begin{equation}
(k,A)_{(B)}  \equiv g^{im (B)} k_i A_m , \quad H^{(\alpha)} \equiv
g^{in (\alpha)} k_i k_n . \label{kA}
\end{equation}
Explicitly we obtain
\begin{equation}
(k,A)_{(B)} = \mp \hat{\mu} \sqrt{\frac{M_{(1)} M_{(2)}
M_{(3)}}{\gamma}} \sum_{(c)= (1) }^{(3)} \frac{\varepsilon_{(c)}
k_{(c)} A_{(c)}}{M_{(c)}} , \label{kA2}
\end{equation}
\begin{equation}
H^{(A)} = (k_i U^i)^2 - \hat{\mu} \sum_{(c)= (1)}^{(3)}
\varepsilon_{(c)} k^2_{(c)},  \label{HA}
\end{equation}
\begin{equation}
H^{(A)} - H^{(B)} = \pm  \hat{\mu} \sqrt{\frac{M_{(1)} M_{(2)}
M_{(3)}}{\gamma}} \sum_{(c)= (1) }^{(3)} \frac{\varepsilon_{(c)}
k^2_{(c)}}{M_{(c)}} . \label{HA-HB}
\end{equation}
The equations (\ref{CA-CB}) may then be written as
\begin{eqnarray}
\delta^{(a)(b)} \left\{ \frac{\hat{\mu}\sqrt{M_{(1)} M_{(2)}
M_{(3)}}}{M_{(a)}} \left[ A_{(b)} \sum_{(c)= (1) }^{(3)}
\frac{\varepsilon_{(c)} k^2_{(c)}}{M_{(c)}}
\right.\right.\qquad\qquad\qquad\qquad\qquad &&\nonumber\\{-}
\left.\left. k_{(b)} \sum_{(c)= (1) }^{(3)}
\frac{\varepsilon_{(c)} k_{(c)} A_{(c)}}{M_{(c)}} \right] {+}
A_{(b)} H^{(A)} \right\} {=} 0 , && \label{system1}
\end{eqnarray}
or, in components,
\begin{eqnarray}
\left( \frac{H^{(A)}}{\hat{\mu} M_{(2)} M_{(3)}} +
\frac{\varepsilon_{(2)} k^2_{(2)}}{M_{(2)}} +
\frac{\varepsilon_{(3)} k^2_{(3)}}{M_{(3)}} \right)A_{(1)} -
\frac{\varepsilon_{(2)} k_{(1)}k_{(2)}}{M_{(2)}}A_{(2)}
\quad &&\nonumber\\
- \frac{\varepsilon_{(3)} k_{(1)}k_{(3)}}{M_{(3)}} A_{(3)}  = 0 ,
\label{system2}
\end{eqnarray}
\begin{eqnarray}
- \frac{\varepsilon_{(1)} k_{(1)}k_{(2)}}{M_{(1)}}A_{(1)}  +
\left( \frac{H^{(A)}}{\hat{\mu} M_{(1)} M_{(3)}} +
\frac{\varepsilon_{(1)} k^2_{(1)}}{M_{(1)}} +
\frac{\varepsilon_{(3)} k^2_{(3)}}{M_{(3)}} \right)A_{(2)} \quad
&&\nonumber\\- \frac{\varepsilon_{(3)}
k_{(2)}k_{(3)}}{M_{(3)}}A_{(3)}  = 0 , \label{system3}
\end{eqnarray}
\begin{eqnarray}
-\frac{\varepsilon_{(1)} k_{(1)}k_{(3)}}{M_{(1)}} A_{(1)}  -
\frac{\varepsilon_{(2)} k_{(2)}k_{(3)}}{M_{(2)}}A_{(2)}
\qquad\qquad\qquad\qquad\qquad \qquad &&\nonumber\\+ \left(
\frac{H^{(A)}}{\hat{\mu} M_{(1)} M_{(2)}} +
\frac{\varepsilon_{(1)} k^2_{(1)}}{M_{(1)}} +
\frac{\varepsilon_{(2)} k^2_{(2)}}{M_{(2)}}  \right)A_{(3)} =0 .&&
\label{system4}
\end{eqnarray}
In addition, we have to consider Eq.~(\ref{LorentzgA}), which in
terms of the tetrad components reads
\begin{equation}
k_{(1)} \varepsilon_{(1)} A_{(1)} + k_{(2)} \varepsilon_{(2)}
A_{(2)} + k_{(3)} \varepsilon_{(3)} A_{(3)} = 0 .
\label{conseqLorentz}
\end{equation}
Formally, we have four equations to find three amplitude
coefficients $A_{(a)}$. However, one can check directly, that one
of the set of equations (\ref{system2}) - (\ref{system4}), say
(\ref{system4}), happens to be a linear combination of two others,
(\ref{system2}) and (\ref{system3}, and of (\ref{conseqLorentz}).
Thus, we can find one of the quantities, say
$k_{(3)}\varepsilon_{(3)} A_{(3)}$, from (\ref{conseqLorentz}) and
insert it into (\ref{system2}) and (\ref{system3}. This provides
us with a system of two equations for $A_{(1)}$ and $A_{(2)}$:
\begin{eqnarray}
\left( \frac{H^{(E)}}{\hat{\mu} M_{(2)} M_{(3)}} {+}
\frac{\varepsilon_{(2)} k^2_{(2)}}{M_{(2)}} {+}
\frac{\varepsilon_{(1)} k^2_{(1)} {+} \varepsilon_{(3)}
k^2_{(3)}}{M_{(3)}} \right) A_{(1)}\qquad\qquad\qquad && \nonumber\\
\qquad\qquad\qquad - \left(\frac{1}{M_{(2)}} {-} \frac{1}{M_{(3)}}
\right) \varepsilon_{(2)} k_{(1)}k_{(2)} A_{(2)} {=} 0 ,
\label{system2eq1}
\end{eqnarray}
\begin{eqnarray}
\left( \frac{1}{M_{(1)}} {-} \frac{1}{M_{(3)}}
\right)\varepsilon_{(1)} k_{(1)}k_{(2)} A_{(1)}
\qquad\qquad\qquad\qquad\qquad\qquad\qquad &&
 \nonumber\\\quad {-} \left( \frac{H^{(E)}}{\hat{\mu} M_{(1)} M_{(3)}}
{+} \frac{\varepsilon_{(1)} k^2_{(1)}}{M_{(1)}} {+}
\frac{\varepsilon_{(2)} k^2_{(2)} {+} \varepsilon_{(3)}
k^2_{(3)}}{M_{(3)}} \right)A_{(2)}  {=} 0 . \label{system2eq2}
\end{eqnarray}
\\
The solution of the system (\ref{system2eq1}), (\ref{system2eq2})
requires the Cramer determinant to be zero. The resulting
characteristic equation is
\\
\begin{eqnarray}
\pm \sqrt{\gamma M_{(1)}M_{(2)}M_{(3)}} \left( \varepsilon_{(1)}
k^2_{(1)} + \varepsilon_{(2)}k^2_{(2)} + \varepsilon_{(3)}
k^2_{(3)}\right) \left( \frac{H^{\left(A\right)} -
H^{\left(B\right)}}{\hat{\mu}} \right)\qquad &&\nonumber\\
+ \left(\frac{H^{(A)}}{\hat{\mu}}\right) \left[ (M_{(2)} +
M_{(3)}) \varepsilon_{(1)} k^2_{(1)} + (M_ {(1)} + M_{(3)})
\varepsilon_{(2)} k^2_{(2)}\right. \qquad && \nonumber\\ +
\left.(M_{(1)} + M_{(2)}) \varepsilon_{(3)} k^2_{(3)}
 + \left(\frac{H^{(A)}}{\hat{\mu}}\right)
 \right] = 0 .
\label{characteq}
\end{eqnarray}
\\
This is a fourth order equation for the wave vector. It
corresponds to the extended Fresnel equation in \cite{HehlObukhov}
(Equations (D.2.23) and (D.2.44)). In the isotropic limit $M_{1} =
M_{2} = M_{3} = 0$ we have the double solution
\\
\begin{equation}
\left(H^{(A)}\right)^{2} = \left(H^{(B)}\right)^{2} =
\left(g^{*ij}k_{i}k_{j}\right)^{2} = 0 , \label{doublesol}
\end{equation}
\\
which coincides with (\ref{nulliso}).  For $n^{2} = 1$ we recover,
of course, the vacuum case (\ref{nullvectorvac}).

\subsection{The uniaxial case}

Obviously, $H^{(A)} = 0$ is also a solution of (\ref{characteq})
if only two of the three quantities $M_{1}$, $M_{2}$ and $M_{3}$
are zero. Let's assume here $M_{2} = M_{3} = 0$. For the uniaxial
configuration
\\
\begin{equation}
\varepsilon_{(2)} = \varepsilon_{(3)} = \varepsilon ,\quad
\varepsilon_{(1)} = \varepsilon\left(1 - \xi\right) ,\quad
\mu_{\left(a\right)} = \mu\ ,\label{conduniax}
\end{equation}
\\
   where $0 < \xi < 1$,    we have
\\
\begin{equation}
M_{2} = M_{3} = 0\quad \Rightarrow\quad \hat{\mu} =
\frac{1}{n^{2}\varepsilon\left(1 - \xi\right)}\quad
\Rightarrow\quad M _{1} = \xi\ .\label{hatmu}
\end{equation}
\\
For the expression multiplying $H^{(A)}/\hat{\mu}$ in the brackets
in (\ref{characteq}) we may write
\\
\begin{eqnarray}
\left(M_{(2)} + M_{(3)}\right) \varepsilon_{(1)} k^2_{(1)} +
\left(M_{(1)} + M_{(3)}\right) \varepsilon_{(2)} k^2_{(2)}
\qquad\qquad\qquad\qquad\quad && \nonumber\\  \quad +
\left(M_{(1)} + M_{(2)}\right) \varepsilon_{(3)} k^2_{(3)} =
\xi\varepsilon\left( k^2_{(2)}
 + k^2_{(3)}\right) .&& \label{}
\end{eqnarray}
\\
On the other hand, from (\ref{HA-HB}),
\\
\begin{equation}
\frac{H^{(A)} - H^{(B)} }{\hat{\mu}} = \pm
\sqrt{\frac{M_1}{\gamma}}\varepsilon\left( k^2_{(2)}
 + k^2_{(3)}\right)\ .\label{HA-HB2}
\end{equation}
\\
With the help of the last relation the characteristic equation
reduces to
\\
\begin{equation}
\frac{H^{(A)}}{\hat{\mu}} \left[\frac{H^{(B)}}{\hat{\mu}} +
\varepsilon\left( k^2_{(2)} + k^2_{(3)}\right)\left(\xi \pm
\sqrt{\frac{\xi}{\gamma}}\right)\right] =   0\ .\label{HAbracket}
\end{equation}
\\
For the choice
\begin{equation}
\gamma = \frac{1}{\xi} \label{gamma=}
\end{equation}
and with the minus sign equations (\ref{HAbracket}) factorizes
into
\\
\begin{equation}
H^{(A)}H^{(B)} =  \left(g^{mn (A)} k_m k_n\right)\left(g^{ij (B)}
k_i k_j\right) = 0\ \label{HAHB=0}
\end{equation}
\\
with solutions $g^{mn (A)} k_m k_n = 0$ and $g^{ij (B)} k_i k_j =
0$. Both the symmetric tensor fields $g^{mn (A)}$ and $g^{ij (B)}$
are optical metrics in such a case, i.e., there are two light
cones describing the phenomenon of birefringence. The first
optical metric is
\\
\begin{equation}
g^{mn (A)} = U^{m}U^{n} - \frac{1}{n^{2}}X^m_{(1)} X^n_{(1)} -
\frac{1}{n^{2}\left(1 - \xi\right)}\left(X^m_{(2)} X^n_{(2)} +
X^m_{(3)} X^n_{(3)}\right)\ .\label{gA}
\end{equation}
\\
The solution $g^{mn (A)} k_m k_n = 0$ corresponds to a dispersion
relation
\\
\begin{equation}
\omega^{2} = \frac{1}{n^{2}}\left(k_{(1)}^{2} + \frac{k_{(2)}^{2}
+ k_{(3)}^{2}}{1 - \xi}\right)\ ,\label{omegaA}
\end{equation}
\\
where $\omega \equiv k_m U^m$. This characterizes the so-called
{\it extraordinary} wave. With the refractive index three-vector
$n_{(a)}$, defined by
\\
\begin{equation}
k_{(a)} = \omega n_{(a)} , \label{defna}
\end{equation}
\\
it can be described as a Fresnel ellipsoid
\\
\begin{equation}
\frac{n_{(1)}^{2}}{n^{2}} + \frac{n_{(2)}^{2}}{n^{2}\left(1 -
\xi\right)} + \frac{n_{(3)}^{2}}{n^{2}\left(1 - \xi\right)} = 1 .
\label{Fresnel}
\end{equation}
\\
In the expression (\ref{gBgeneral}) for $g^{mn (B)}$ the
anisotropic contributions exactly cancel for $\gamma = \xi^{-1}$
and we recover
\\
\begin{equation}
g^{mn (B)} = U^{m}U^{n} + \frac{1}{n^{2}}\Delta^{mn}\ ,\label{gB}
\end{equation}
\\
which coincides with the optical metric $g^{*mn}$ in
(\ref{gstar}).  Here, the dispersion relation is
\\
\begin{equation}
\omega^{2} = \frac{1}{n^{2}}\left(k_{(1)}^{2} + k_{(2)}^{2} +
k_{(3)}^{2}\right)\ , \label{omegaB}
\end{equation}
\\
which describes {\it ordinary} wave propagation.

Consequently, we have obtained an explicit representation
(\ref{maindecomp}) for the material tensor $C^{ikmn}$ with the
optical metrics (\ref{gA}) and (\ref{gB}), with $\hat{\mu}$ from
(\ref{hatmu}) and with $\gamma = \xi^{-1}$ according to
(\ref{gamma=}). This explicit representation within the context of
the geometric background revealed in (\ref{trafo1AB}) -
(\ref{determinants}) is the main result of this paper. While the
general form of the decomposition (\ref{maindecomp}) remains true
in the general case, the interpretation of $g^{mn (A)}$ and $g^{mn
(B)}$ as optical metrics is valid only for uniaxial symmetry. Of
course, the relations (\ref{trafo1AB}) - (\ref{trafoB12}) between
the optical metrics $g^{mn (A)}$ and $g^{mn (B)}$ and $g^{mn (1)}$
and $g^{mn (2)}$ hold in this special case as well. This means, a
representation (\ref{general}) is possible although $g^{mn (1)}$
and $g^{mn (2)}$ do not have an obvious physical meaning.

\subsection{The energy of an electromagnetic wave}

Finally, we briefly discuss some features of the electromagnetic
energy-momentum tensor in this approach. The Minkowski tensor of
the electromagnetic field reads
\begin{equation}
T_{ij} \equiv  \frac{1}{4} g_{ij} F_{mn} H^{mn} - F_{im} H_{jn}
g^{mn} \label{mink}
\end{equation}
Taking into account Maxwell's equations (\ref{Maxgo}) and the
representation (\ref{Fmn}) of the field strength tensor (notice
that for the square of the field strength the complex conjugate
has to be taken), one realizes that the scalar $F_{mn} H^{mn}$
vanishes and the energy-momentum tensor takes the simple form
\begin{equation}
T_{ij} = -  2 k_i C_j^{\cdot lmn} A_l k_m A_n. \label{mink1}
\end{equation}
In terms of the optical metrics and with (\ref{LorentzgA}) and
(\ref{ConditionsG}) the energy scalar is equal to
\begin{equation}
W \equiv T_{ij} U^i U^j = - \frac{\omega^2}{\hat{\mu}} g^{ln(A)}
A_l A_n = - \omega^2\varepsilon^{ln} A_l A_n . \label{mink31}
\end{equation}
\\
With (\ref{eigen1}), (\ref{exx}) and (\ref{decompkA}) one checks
that this quantity is positive. Incidentally, the energy density
of the electromagnetic field does not depend on whether it is
calculated from the Minkowski or from the Abraham tensor
\cite{Brevik1}. Remarkably, it is only the extraordinary optical
metric $g^{ln(A)}$ from (\ref{gA}) which enters the expression
(\ref{mink31}). The ordinary optical metric $g^{ln(B)}$ does not
contribute here.

Generally, the Minkowski tensor is not symmetric. In the isotropic
case, however, there exists a generalized symmetry (cf.
\cite{HehlObukhov}, (E.3.50)). With (\ref{mink1}),
(\ref{Cisogstar}), (\ref{invers}) and (\ref{Lorentziso})  it takes
the form
\begin{equation}
g^{*}_{ia}T^{i}_{\,\, b} =
\frac{n^{4}}{\mu}k_{b}k_{a}g^{*kn}A_{k}A_{n} =
g^{*}_{ib}T^{i}_{\,\, a} \label{}
\end{equation}
in the present context. A corresponding relation is not found in
the anisotropic case.

\section{Discussion}
\label{Discussion}

We have shown that the material tensor of an anisotropic medium
with uniaxial symmetry can be constructed out of those two optical
metrics which describe ordinary and extraordinary light
propagation. This generalizes the corresponding representation of
the material tensor in isotropic media in terms of Gordon's
optical metric. Furthermore, we have clarified the internal
geometrical structure which underlies the decomposition of the
material tensor into a combination of products of two symmetric
second rank tensor fields. The latter quantities can be regarded
as vectors in an associated two dimensional space where the
coefficients of the decomposition play the role of a metric in
this space. While we believe this type of bi-metric representation
in connection with the interpretation in terms of an internal
metric structure to be new, some already known results are
reproduced here in a different context as well. Generally, the
wave propagation in anisotropic media is governed by a quartic
equation for the wave vector, called (extended) Fresnel equation
\cite{HehlObukhov}. This equation is known to factorize into two
quadratic equations in vacuum. A similar feature is true for
isotropic media. Here, the factorizing implies a double solution
for the wave vector as a null vector with respect to the optical
metric. For an anisotropic medium with uniaxial symmetry a
factorization is obtained as well. Instead of the isotropic double
solution there are two solutions in terms of two optical metrics
now, one of them describing ordinary, the other one extraordinary
wave propagation. All these features follow as special cases of
our approach. Furthermore, we recover a (generalized) symmetry
property of Minkowski's energy-momentum tensor for isotropic
media.

\section*{Acknowledgments}
This work was supported by the Deutsche Forschungsgemeinschaft.

\end{document}